\documentclass[prl,aps,twocolumn,floatfix,superscriptaddress,a4paper,showpacs]{revtex4}

\usepackage{amsmath,amssymb,amsfonts}
\usepackage{bm}
\usepackage{graphicx}

\usepackage{color}

\begin{document}

\title{Spin dynamics in a one-dimensional ferromagnetic Bose gas}

\author{M. B. Zvonarev}
\affiliation{DPMC-MaNEP, University of Geneva, 24 quai Ernest-Ansermet, 1211 Geneva 4, Switzerland}

\author{V. V. Cheianov}
\affiliation{Physics Department, Lancaster University, Lancaster, LA1 4YB, UK}

\author{T. Giamarchi}
\affiliation{DPMC-MaNEP, University of Geneva, 24 quai Ernest-Ansermet, 1211 Geneva 4, Switzerland}

\begin{abstract}
We investigate the propagation of spin excitations in a one-dimensional (1D)
ferromagnetic Bose gas. While the spectrum of longitudinal spin
waves in this system is sound-like, the dispersion of
transverse spin excitations is quadratic making a direct application
of the Luttinger Liquid (LL) theory impossible. By using a combination
of different analytic methods we derive the large time asymptotic behavior
of the spin-spin dynamical correlation function for strong interparticle
repulsion. The result has an unusual structure associated with a crossover from the regime of trapped spin wave to an open regime and does not have analogues in known low-energy universality classes of quantum 1D systems.

\end{abstract}

\date{\today}
\maketitle

Quantum interacting one-dimensional (1D) systems have for many decades been a subject of a ceaseless interest of both theorists and experimentalists. This is mainly because of the unique role of quantum fluctuations, which are so strong in 1D that even for weakly interacting systems the intuition based on the free-particle picture and the mean field theory fails and the effects of strong correlations become important~\cite{giamarchi_book_1d}. Such effects have been encountered in many experiments dealing with 1D conductors, like, {\it e.g.}, organic salts, quantum wires or carbon nanotubes, in which the constituent particles, electrons, are spin 1/2 fermions.

Recent advances in the creation and manipulation of ultracold atomic gases~\cite{bloch_QG_review} opened an access to a new class of 1D systems where the constituent particles obey bosonic statistics~\cite{stoferle_tonks_optical,paredes_tonks_optical, kinoshita_tonks_continuous, Kinoshita_1DBosons_g2, Tolra_1DBosons_g3} and have a variable number of internal (``spin'') states~\cite{RB87_experiment_spin_waves_resolution_Gurik,RB87_experiment_magnetization_imaging_Higbie,Sadler_1DBosons_RB87_ferromagnet_breaking}.
In the spinless case the theory predicts an equivalence between the Bose and Fermi systems: both are described by the Luttinger Liquid (LL) theory at low energies~\cite{giamarchi_book_1d}. In the presence of spin the situation is more complex. The necessary condition for the applicability of the LL theory is the linearity of the dispersion of low-lying elementary excitations, $\varepsilon(p)\sim\vert p\vert.$ This is usually the case for fermions, which have a natural tendency to anti-ferromagnetic ordering~\footnote{Lieb and Mattis showed rigorously~\cite{lieb_mattis_fermions_unmagnetized} that for the spin-independent interaction the ground state of 1D fermions is completely unpolarized.}. While under special conditions linear dispersion relations, and, consequently, the LL physics can be encountered in a multi-component Bose system~\cite{kleine_bosons_spincharge}, there exists a broad range of Hamiltonians, in particular those with spin-independent interactions, whose ground state is ferromagnetic~\cite{eisenberg_polarization_bosons_spin}. In the latter case the softest low-lying excitation is the magnon with a quadratic dispersion relation
\begin{equation}
\varepsilon(p)\simeq p^2/2m_*, \qquad p\to0, \label{dispersion}
\end{equation}
where $m_*$ is an effective mass. This makes the straightforward application of the LL theory impossible and poses a fundamental question of finding an alternative theory describing the dynamics of the low-energy excitations in a 1D boson ferromagnet.

In this Letter we tackle this issue and compute the long-distance properties of two-point correlation functions of local spins at zero temperature. Focusing on the region of strong interparticle repulsion, we show that the dynamical properties of spin excitations in a 1D ferromagnetic Bose gas are neither those of a localized ferromagnet nor of a Luttinger liquid, pointing at the existence of a new low-energy universality class. Our main results are presented in Eqs.~\eqref{answer} and \eqref{alphabeta}. We also discuss the connection of our work with the problems of a moving impurity in a LL, dynamics of a hole in the Hubbard-Mott insulator, and quantum mechanics in a dissipative environment. Finally, we describe recent experimental realizations~\cite{RB87_experiment_spin_waves_resolution_Gurik, RB87_experiment_magnetization_imaging_Higbie, Sadler_1DBosons_RB87_ferromagnet_breaking} of quasi-1D Bose gases with spin.

For simplicity of presentation the derivations are carried out for two-component bosons; the generalization to higher spins is straightforward~\cite{zvonarev_ferrobosons_long}. We assume that the interaction between the particles is spin-independent and the model Hamiltonian has the form
\begin{equation}
H= \sum_{j=1}^N \frac{p_j^2}{2m}+ \sum_{i<j} [g\delta(x_i-x_j)+ U(x_i-x_j)]+ h S_z. \label{Ham1}
\end{equation}
Here $m$ is the mass of a boson,  $h$ is the external magnetic field, and $S_z$ is the $z$ component of the total spin. We are interested in the limit of infinite number of particles, $N\to\infty,$ and of infinite system size, $L\to \infty,$ at a fixed particle density, $\rho_0=N/L.$ Although for cold atoms the interaction potential is well approximated by a delta-function, we allow for a more general interaction $g \delta(x)+U(x),$ where $U(x)$ is some smooth function. The strength of the short-range repulsion is characterized by a dimensionless coupling constant $\gamma=mg/{\hbar^2 \rho_0}.$ For $U=0$ the Hamiltonian \eqref{Ham1} can be diagonalized by Bethe Ansatz (BA)~\cite{gaudin_book}, providing us with a valuable source of intuition about the low-energy dynamics studied here.

The global spin operator $\mathbf{S}=(S_x,S_y,S_z)$ can be represented as
\begin{equation}
\mathbf{S}=\int_0^L dx\, \mathbf{s}(x), \quad \mathbf{s}(x)=\frac{1}{2}  \sum_{j=1}^N \bm \sigma^{(j)}\, \delta(x-x_j),
\end{equation}
where $\bm \sigma^{(j)}=(\sigma_x, \sigma_y, \sigma_z) $ is the vector composed of the three Pauli matrices acting non-trivially on the
spin indices of the $j$-th particle. Spin-ladder operators $s_{\pm}(x)=s_x(x)\pm i s_y(x)$ flip the $z$ component of a local spin. For $h>0$ the Hamiltonian \eqref{Ham1} has a non-degenerate ground state, $|\Uparrow\rangle,$ which is fully polarized along the $z$-axis, $s_{+}(x)|\Uparrow\rangle=0.$ The degeneracy appears at $h=0$ and is discussed in detail in Ref.~\cite{eisenberg_polarization_bosons_spin}. We investigate the dynamics of excitations over the state $|\Uparrow\rangle.$ For simplicity of the presentation, our results are given for $h=0.$ In the case of $h\ne0$ the r.h.s.\ of Eqs.~\eqref{G_perp2} and~\eqref{answer} should be multiplied by the oscillating factor $e^{ith}.$

\begin{figure}
\begin{center}
\includegraphics[clip,width=8cm]{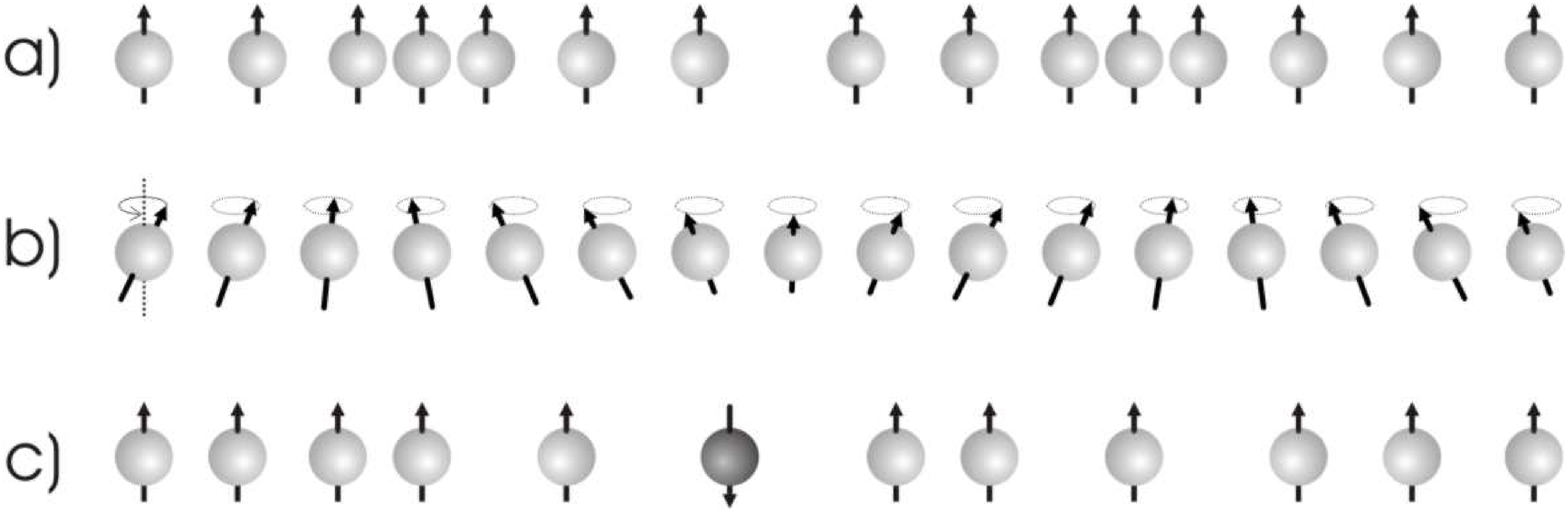}
\end{center}
\caption{A 1D array of particles carrying spin. The propagation of longitudinal (a) and transverse (b) spin waves over the fully polarized state $|\Uparrow\rangle$ is depicted. The state obtained by the action of the spin lowering operator $s_{-}(0,0)$ onto $|\Uparrow\rangle$ is schematically illustrated in (c). } \label{fig_spin}
\end{figure}

Our main object of interest is the response of the system to weak perturbations of local magnetization, encoded into two correlation functions:
\begin{align}
& G_{\parallel}(x,t)=\langle\Uparrow| s_z(x,t)s_z(0,0) |\Uparrow\rangle \label{G_par} \\
& G_{\perp}(x,t)=\langle\Uparrow| s_+(x,t)s_-(0,0) |\Uparrow\rangle \label{G_perp}
\end{align}
describing the longitudinal, Fig.~\ref{fig_spin}(a), and transverse, Fig.~\ref{fig_spin}(b), spin dynamics, respectively. The longitudinal dynamics over the state $|\Uparrow\rangle$ coincides with that of density fluctuations in the 1D spinless Bose gas. Their dispersion is linear in the low-energy limit, therefore the $x,t\to\infty$ asymptotics of Eq.~\eqref{G_par} can be calculated within the LL theory \cite{giamarchi_book_1d}, and decays as a power-law~\footnote{The asymptotic expression for $G_\parallel$ can be found, for example, in Ref.~\cite{giamarchi_book_1d}, Eq.~(11.7).}.

In contrast to the longitudinal dynamics, transverse spin waves are not sound waves, their dispersion is quadratic at low momenta, Eq.~\eqref{dispersion}.
The effective mass $m_*$ in Eq.~\eqref{dispersion} increases with increasing $\gamma.$ In the BA solvable case $m_*$ can be calculated exactly \cite{fuchs_spin_waves_1D_bose} and shows a linear divergence with $\gamma$ in the limit of strong repulsion: $m_*/m\simeq 3\gamma/2\pi^2$ as $\gamma\to\infty.$ This divergency was pointed out in Ref.~\cite{fuchs_spin_waves_1D_bose} as a signature of a slow dynamics of transverse spin waves at large $\gamma$. We show here that in the strong coupling regime the effects are even more dramatic. To see this, we start with the description of the long-wavelength limit of Eq.~\eqref{G_perp} at $\gamma=\infty$.

Qualitatively, the propagation of the transverse spin excitation at $\gamma=\infty$ can be described as follows: the operator $s_-(0,0)$ flips a spin of a given particle, shown in Fig.~\ref{fig_spin}(c) with a down arrow. Because of the infinite repulsion, the spin-down particle cannot exchange its position with its neighbors, so it is trapped inside an interval of average length $2\rho_0^{-1}$. However, since the spin-up particles are mobile, the size and the position of this interval can fluctuate, making it possible for the spin-down particle to be detected at a distance larger than $\rho_0^{-1}$ from its initial position. Such fluctuations require a simultaneous displacement of a large number of spin-up particles, thus the probability to observe the spin-down particle at a large distance from the point where it was created must be small. The correlation function \eqref{G_perp} should thus decay very rapidly with distance.

To quantify the above argument we obtain the analytic expression for the long-wavelength asymptotics of Eq.~\eqref{G_perp}. The calculations are performed using, as in Ref.~\cite{fiete_spin_decoherent}, a combination of bosonization with a first-quantized path integral. For the BA solvable case they can be underpinned by the asymptotic analysis of the determinant representation of the correlation functions along the lines of Refs.~\cite{cheianov_spin_decoherent_short,cheianov_spin_decoherent_long}. We give only the result here, the details will be presented elsewhere \cite{zvonarev_ferrobosons_long}. The asymptotic form of $G_\perp (x,t)$ is given for $t\gg t_\mathrm{F}$ by:
\begin{equation}
G_\perp (x,t)\simeq \frac1{\sqrt{\ln (t/t_\mathrm{F})}}
\exp\left\{-\frac1K \frac{(\pi\rho_0 x)^2}{2\ln (t/t_\mathrm{F})}\right\}.
\label{G_perp2}
\end{equation}
It is similar to the diffusion propagator except that the mean square deviation grows logarithmically with time. This logarithmic diffusion is the mathematical manifestation of the spin trapping effect. The parameter $t_\mathrm{F}$ controls the smallest time scale in the problem:
\begin{equation}
t_\mathrm{F}=\frac{\hbar}{E_\mathrm{F}}, \quad E_\mathrm{F}= \frac{\hbar^2}{2m}(\pi\rho_0)^2. \label{t_0}
\end{equation}
The physical meaning of $t_\mathrm{F}$ can be understood from the fact that the spectrum of the $\delta$-interacting spinless bosons at $\gamma=\infty$ is the same as that of free spinless fermions. Therefore, $E_\mathrm{F}$ in Eq.~\eqref{t_0} plays the role of a Fermi energy, and $t_\mathrm{F}$ is the time scale for the longitudinal spin fluctuations. The dimensionless parameter $K$ in Eq.~\eqref{G_perp2} is the Luttinger parameter, which can be calculated from the thermodynamic properties of the system~\cite{giamarchi_book_1d}. Note that for $U=0$ one has $K\ge1$ and $K\to 1$ only at $\gamma\to\infty$.

Eq.~\eqref{G_perp2} is obtained for $\gamma=\infty$. For large but finite $\gamma$ there is a small probability that spin-up and spin-down particles exchange their positions, allowing the spin excitation to escape from the trap. We estimate the escape time $t_*$ by replacing the fluctuating gas of spin-up particles by a static Kronig-Penney lattice with period $\rho_0^{-1}$ and get $t_*\sim \gamma t_\mathrm{F}.$ Thus, for large $\gamma$ the escape time is parametrically large and there exists a broad window $t_\mathrm{F} \ll t \ll t_*$ where $G_\perp (x,t)$ has the asymptotics form~\eqref{G_perp2}.

For $t\gtrsim t_*,$ there is a crossover to another, ``open'', regime, which we investigate next. A rather general result can be obtained from a minimal set of assumptions on the analytic properties of the spectral function $A(k,\omega).$ By definition:
\begin{equation}
G_{\perp} (x,t) = \int \frac{dk}{2\pi} e^{i k x} \int \frac{d\omega}{2\pi} e^{-i \omega t} A(k, \omega)
\label{specrep}
\end{equation}
with
\begin{equation}
A(k, \omega)=\sum_{\nu} \delta(\hbar\omega-E_{\nu}(k)) \vert \langle \nu, k \vert s_{-}(k)\vert \rm \Uparrow \rangle\vert^2.
\label{specrep2}
\end{equation}
Here $H|\nu,k\rangle= E_\nu(k)|\nu,k\rangle$ and $\nu$ enumerates all the states with a given momentum $\hbar k.$ Among these states there is a state of a minimal energy $\varepsilon(k)$, that is $\varepsilon(k)\equiv\min_\nu E_\nu(k).$ Our first assumption, supported by the calculations for the BA solvable case~\cite{zvonarev_ferrobosons_long}, where $\varepsilon(k)$ can be found explicitly, and variational considerations in the spirit of the Feynmann single-mode approximation is that at small $p=\hbar k$ the function $\varepsilon(k)$ has the form~\eqref{dispersion}. Therefore, $\varepsilon(k)$ defines the threshold frequency for the spectral function: $A(k,\omega)=0$ for $\hbar\omega<\varepsilon(k).$ Above the threshold a continuum of states contributes to $A(k,\omega)$. They contain one magnon (spin-flip) and multiple plasmon (density) excitations close to zero momentum. The large $t$ asymptotics of Eq.~\eqref{specrep} is dominated by the scale-free part of the spectral function at the threshold, whose most generic form is
\begin{equation}
A(k,\omega)\simeq c(k) [\hbar\omega-\varepsilon(k)]^{\Delta(k)}, \quad \hbar\omega\ge\varepsilon(k),
\label{Athresh}
\end{equation}
where $\Delta(k)$ and $c(k)$ are some functions of $k$.

Our second assumption is that both $c(k)$ and $\Delta(k)$ are analytic functions of momentum in the vicinity of $k=0.$ Taking into account the inversion symmetry, $G_\perp(x,t)=G_\perp(-x,t),$ which implies $A(k,\omega)=A(-k,\omega),$ the general form of $c(k)$ and $\Delta(k)$ should be
\begin{align}
& \Delta(k)=\alpha-1+\beta k^2+\cdots, \label{Delta}\\
& c(k)=c_0+c_1 k^2+\cdots,
\label{c}
\end{align}
where $\alpha,$ $\beta,$ $c_0$ and $c_1$ are some model-dependent coefficients. We substitute Eqs.~\eqref{Athresh}, \eqref{Delta} and \eqref{c} into Eq.~\eqref{specrep}, and for the function $\varepsilon(k)$ entering Eq.~\eqref{Athresh} we use formula \eqref{dispersion} with $p=\hbar k$. The saddle point analysis of the resulting expression leads to our main result:
\begin{multline}
G_{\perp} (x,t)\simeq
t^{-\alpha} \left[\beta \ln\left(\frac{t}{t_\mathrm{F}}\right)+\frac{it\hbar}{2m_*}\right]^{-1/2}\\
\times\exp\left\{\frac{i m_* x^2}{2t\hbar-4 i \beta m_* \ln(t/t_\mathrm{F})}\right\}.
\label{answer}
\end{multline}
Eq.~\eqref{answer} contains two parameters, $\alpha$ and $\beta,$ which depend on both $\gamma$ and $U.$ This dependence cannot be extracted from the scaling and analyticity assumptions, Eqs.~\eqref{Athresh} and \eqref{Delta}. However, we can find $\alpha$ and $\beta$ in the trapped regime, $\gamma=\infty$, by letting $m_*=\infty$ in Eq.~\eqref{answer} and comparing the resulting expression with Eq.~\eqref{G_perp2}. We get
\begin{equation}
\alpha= 0, \qquad \beta= \frac{K}{2 (\pi\rho_0)^2}
\label{alphabeta}
\end{equation}
at $\gamma= \infty$. In addition, for $U=0$ we could treat the case of large but finite $\gamma$ by BA and showed that Eq.~\eqref{alphabeta} remains valid \cite{zvonarev_ferrobosons_long}. Whether this is an artefact of integrability or a generic property of the model is an important open question. The latter would mean that the asymptotic behavior of the correlation functions in the ferromagnetic case is completely determined by thermodynamic properties of the system, like in the LL theory.

If $m_*<\infty$ equation~\eqref{answer} coincides with \eqref{G_perp2} for sufficiently small times, $2t\hbar\ll 4\beta m_* \ln(t/t_\mathrm{F}).$ At a time $t_*$ when this relation becomes an equality a crossover to an open regime occurs~\footnote{The crossover between these two regimes corresponds in $A(k,\omega)$ to change of sign of the $k^2$ term
in a small $k$ expansion. Its physics is however seen much more transparently in real space and time.}. For $U=0$ the dependence of $t_*$ on $\gamma$ can be found explicitly:
\begin{equation}
t_*\simeq \frac{3\gamma K}{4\pi^2} \ln\left(\frac{3\gamma K}{4\pi^2} \right) t_\mathrm{F}. \label{t_*}
\end{equation}
We stress that this estimate is valid if $\gamma$ is large enough to ensure $t_*\gg t_\mathrm{F}$. This confirms and completes our naive estimate of $t_*$ given in the paragraph below Eq.~\eqref{G_perp2}.

\begin{figure}
\begin{center}
\includegraphics[clip,width=8cm]{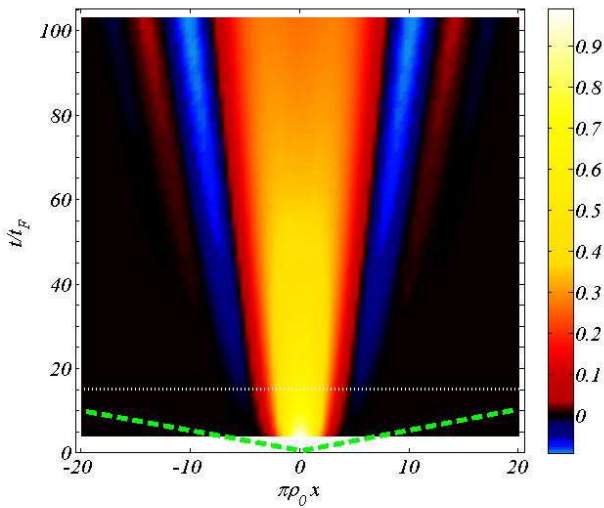}
\end{center}
\caption{(color online) Shown is the intensity plot of $t^\alpha\mathrm{Re}\,G_\perp(x,t)$ in the ${x,t}$ plane at $\gamma=100$. The spin wave is strongly suppressed at a distance $x\sim \ell(t)$, Eq.~\eqref{ell}, long before it reaches the light cone, $x=vt,$ shown as dashed green lines. The onset of the spin precession is seen in the space oscillations of $t^\alpha\mathrm{Re}\,G_\perp$ developing above the crossover time $t_*\simeq15 t_\mathrm{F},$ shown by the white dotted line.} \label{fig}
\end{figure}

In the open regime, $t \gtrsim t_*,$ the function~\eqref{answer} factorizes into a product of the transverse spin correlation function of the localized Heisenberg ferromagnet, $G_\perp^H,$ exhibiting exchange-induced oscillations, and a rapidly decaying factor, which we attribute to the excitation of the charge degrees of freedom:
\begin{equation}
G_\perp\simeq e^{-\frac{x^2}{2\ell^2}}t^{-\alpha} G_\perp^{H} , \quad \ell(t)=\frac{2K^{-\frac12}}{\pi\rho_0}\frac{t/t_\mathrm{F}}{\sqrt{\ln t/t_\mathrm{F}}} \frac{m}{m_*}.
\label{ell}
\end{equation}
The strong suppression of $G_\perp$ for $x>\ell(t)$ ensures the absence of the light-cone singularities at $x=vt,$ where $v$ is the sound velocity for the charge excitations, $v=\hbar \pi \rho_0/m$ for $\gamma=\infty.$ This is illustrated in Fig.~\ref{fig}.

We now discuss the relation of our work to several previously studied problems. The threshold singularity, Eq.~\eqref{Athresh}, of $A(k,\omega)$ is identical to the Fermi edge singularity in a 1D quantum fluid with a mobile impurity~\cite{castella_mob_impurity93,ogawa_mob_impurity92,tsukamoto_mob_impurity98} in a special case of equal impurity and host particles masses. Our results, therefore, have interesting implications for the latter problem. In particular, the relation of $\Delta(k)$ to the parameters of the model has only been understood at $k=0$~\cite{castella_mob_impurity93,ogawa_mob_impurity92}. The finite $k$ case was addressed in Ref.~\cite{tsukamoto_mob_impurity98} by using an effective field theory. We, however, found that $\Delta(k)$ calculated from the Ref.~\cite{tsukamoto_mob_impurity98} disagrees with our Eqs.~\eqref{Delta} and \eqref{alphabeta}. In particular, it does not reproduce Eq.~\eqref{G_perp2}, which we obtain by two independent methods~\footnote{The results of Ref.~\cite{tsukamoto_mob_impurity98} are largely based on an incorrect BA solution of the mobile impurity problem by Li and Ma (their Ref.~[25]).}. There is also an analogy with the problem of a mobile hole in a 1D Hubbard-Mott insulator~\cite{sorella_hubbmott_mobhole96}. There the parabolic branch of spectrum is due to holon excitations, while the gapless mode with a linear dispersion corresponds to spinons. The spectral function given by  Eq.~(3) of Ref.~\cite{sorella_hubbmott_mobhole96} shows the same critical properties as $A(k,\omega),$ Eqs.~\eqref{Athresh} and \eqref{Delta}. We thus hope that the results presented in this Letter can be extended to a broader class of models. Finally, by bosonizing the gas of spin-up particles one maps the Hamiltonian~\eqref{Ham1} onto that of a quantum particle non-linearly coupled to a harmonic environment. A linearized version of this problem was studied in Refs.~\cite{caldeira_leggett,hakim_dissip85}, and the comparison of the long-time asymptotics remains an open question.

Experiments on quasi-1D Bose gases with internal degrees of freedom
started recently \cite{RB87_experiment_spin_waves_resolution_Gurik,
RB87_experiment_magnetization_imaging_Higbie,
Sadler_1DBosons_RB87_ferromagnet_breaking}. In the
experiment~\cite{RB87_experiment_spin_waves_resolution_Gurik} a
pseudospin $1/2$ system was created by loading $ ^{87}\mathrm{Rb}$
atoms into a highly elongated magnetic trap and selecting two
hyperfine states, $|F=1,m_F=-1\rangle$ and $|F=2,m_F=1\rangle,$
while in \cite{RB87_experiment_magnetization_imaging_Higbie,
Sadler_1DBosons_RB87_ferromagnet_breaking} the true spin $1$ system
was created by loading $ ^{87}\mathrm{Rb}$ atoms into a highly
elongated optical trap and selecting the states composing $F=1$ spin
triplet. In both \cite{RB87_experiment_spin_waves_resolution_Gurik}
and \cite{RB87_experiment_magnetization_imaging_Higbie,
Sadler_1DBosons_RB87_ferromagnet_breaking} long-lived fully
polarized states were achieved and a possibility to excite and image
(pseudo)spin waves in real space and time was demonstrated. Although
in the above experiments the systems were not truly 1D (about $50$
bands of transverse quantization were populated) a possibility to
further reduce the number of occupied bands is suggested by the
successes in the creation of 1D spinless
systems~\cite{stoferle_tonks_optical,kinoshita_tonks_continuous,
Kinoshita_1DBosons_g2,paredes_tonks_optical, Tolra_1DBosons_g3},
thus paving the way to the investigations of the ferromagnetism in 1D Bose gases.

The authors are grateful to M.~Cazalilla, B.~Dou\c{c}ot and Y.~Tserkovnyak for helpful discussions. This work was supported in part by the Swiss NSF under MaNEP and division II and the IHP-Centre Emile Borel.


\begin{thebibliography}{27}
\expandafter\ifx\csname natexlab\endcsname\relax\def\natexlab#1{#1}\fi
\expandafter\ifx\csname bibnamefont\endcsname\relax
  \def\bibnamefont#1{#1}\fi
\expandafter\ifx\csname bibfnamefont\endcsname\relax
  \def\bibfnamefont#1{#1}\fi
\expandafter\ifx\csname citenamefont\endcsname\relax
  \def\citenamefont#1{#1}\fi
\expandafter\ifx\csname url\endcsname\relax
  \def\url#1{\texttt{#1}}\fi
\expandafter\ifx\csname urlprefix\endcsname\relax\def\urlprefix{URL }\fi
\providecommand{\bibinfo}[2]{#2}
\providecommand{\eprint}[2][]{\url{#2}}

\bibitem[{\citenamefont{Giamarchi}(2004)}]{giamarchi_book_1d}
\bibinfo{author}{\bibfnamefont{T.}~\bibnamefont{Giamarchi}},
  \emph{\bibinfo{title}{Quantum Physics in One Dimension}}
  (\bibinfo{publisher}{Oxford University Press}, \bibinfo{address}{Oxford},
  \bibinfo{year}{2004}).

\bibitem[{\citenamefont{Bloch et~al.}(2007)\citenamefont{Bloch, Dalibard, and
  Zwerger}}]{bloch_QG_review}
\bibinfo{author}{\bibfnamefont{I.}~\bibnamefont{Bloch}},
  \bibinfo{author}{\bibfnamefont{J.}~\bibnamefont{Dalibard}}, \bibnamefont{and}
  \bibinfo{author}{\bibfnamefont{W.}~\bibnamefont{Zwerger}},
  \bibinfo{journal}{e-print arXiv:0704.3011}  (\bibinfo{year}{2007}).

\bibitem[{\citenamefont{St{\"o}ferle et~al.}(2004)\citenamefont{St{\"o}ferle,
  Moritz, Schori, K{\"o}hl, and Esslinger}}]{stoferle_tonks_optical}
\bibinfo{author}{\bibfnamefont{T.}~\bibnamefont{St{\"o}ferle} {\it et.~al.}},
  \bibinfo{journal}{Phys. Rev. Lett.} \textbf{\bibinfo{volume}{92}},
  \bibinfo{pages}{130403} (\bibinfo{year}{2004}).

\bibitem[{\citenamefont{Paredes et~al.}(2004)\citenamefont{Paredes, Widera,
  Murg, Mandel, F\"olling, Cirac, Shlyapnikov, H\"ansch, and
  Bloch}}]{paredes_tonks_optical}
\bibinfo{author}{\bibfnamefont{B.}~\bibnamefont{Paredes} {\it et.~al.}},
  \bibinfo{journal}{Nature} \textbf{\bibinfo{volume}{429}},
  \bibinfo{pages}{277} (\bibinfo{year}{2004}).

\bibitem[{\citenamefont{Kinoshita et~al.}(2004)\citenamefont{Kinoshita, Wenger,
  and Weiss}}]{kinoshita_tonks_continuous}
\bibinfo{author}{\bibfnamefont{T.}~\bibnamefont{Kinoshita}},
  \bibinfo{author}{\bibfnamefont{T.}~\bibnamefont{Wenger}}, \bibnamefont{and}
  \bibinfo{author}{\bibfnamefont{D.~S.} \bibnamefont{Weiss}},
  \bibinfo{journal}{Science} \textbf{\bibinfo{volume}{305}},
  \bibinfo{pages}{1125} (\bibinfo{year}{2004}).

\bibitem[{\citenamefont{Kinoshita et~al.}(2005)\citenamefont{Kinoshita, Wenger,
  and Weiss}}]{Kinoshita_1DBosons_g2}
\bibinfo{author}{\bibfnamefont{T.}~\bibnamefont{Kinoshita}},
  \bibinfo{author}{\bibfnamefont{T.}~\bibnamefont{Wenger}}, \bibnamefont{and}
  \bibinfo{author}{\bibfnamefont{D.~S.} \bibnamefont{Weiss}},
  \bibinfo{journal}{Phys. Rev. Lett.} \textbf{\bibinfo{volume}{95}},
  \bibinfo{pages}{190406} (\bibinfo{year}{2005}).

\bibitem[{\citenamefont{Tolra et~al.}(2004)\citenamefont{Tolra, O'Hara,
  Huckans, Phillips, Rolston, and Porto}}]{Tolra_1DBosons_g3}
\bibinfo{author}{\bibfnamefont{B.~L.} \bibnamefont{Tolra} {\it et.~al.}},
  \bibinfo{journal}{Phys. Rev. Lett.} \textbf{\bibinfo{volume}{92}},
  \bibinfo{pages}{190401} (\bibinfo{year}{2004}).

\bibitem[{\citenamefont{McGuirk et~al.}(2002)\citenamefont{McGuirk,
  Lewandowski, Harber, Nikuni, Williams, and
  Cornell}}]{RB87_experiment_spin_waves_resolution_Gurik}
\bibinfo{author}{\bibfnamefont{J.~M.} \bibnamefont{McGuirk} {\it et.~al.}},
 \bibinfo{journal}{Phys. Rev. Lett.}
  \textbf{\bibinfo{volume}{89}}, \bibinfo{pages}{090402}
  (\bibinfo{year}{2002}).

\bibitem[{\citenamefont{Higbie et~al.}(2005)\citenamefont{Higbie, Sadler,
  Inouye, Chikkatur, Leslie, Moore, Savalli, and
  Stamper-Kurn}}]{RB87_experiment_magnetization_imaging_Higbie}
\bibinfo{author}{\bibfnamefont{J.~M.} \bibnamefont{Higbie} {\it et.~al.}},
  \bibinfo{journal}{Phys. Rev. Lett.} \textbf{\bibinfo{volume}{95}},
  \bibinfo{pages}{050401} (\bibinfo{year}{2005}).

\bibitem[{\citenamefont{Sadler et~al.}(2006)\citenamefont{Sadler, Higbie,
  Leslie, Vengalattore, and
  Stamper-Kurn}}]{Sadler_1DBosons_RB87_ferromagnet_breaking}
\bibinfo{author}{\bibfnamefont{L.~E.} \bibnamefont{Sadler} {\it et.~al.}},
 \bibinfo{journal}{Nature}
  \textbf{\bibinfo{volume}{443}}, \bibinfo{pages}{312} (\bibinfo{year}{2006}).

\bibitem[{\citenamefont{Kleine et~al.}(2007)\citenamefont{Kleine, Kollath,
  McCulloch, Giamarchi, and Schollwoeck}}]{kleine_bosons_spincharge}
\bibinfo{author}{\bibfnamefont{A.}~\bibnamefont{Kleine} {\it et.~al.}},
  \bibinfo{journal}{e-print arXiv:0706.0709}  (\bibinfo{year}{2007}).

\bibitem[{\citenamefont{Eisenberg and
  Lieb}(2002)}]{eisenberg_polarization_bosons_spin}
\bibinfo{author}{\bibfnamefont{E.}~\bibnamefont{Eisenberg}} \bibnamefont{and}
  \bibinfo{author}{\bibfnamefont{E.~H.} \bibnamefont{Lieb}},
  \bibinfo{journal}{Phys. Rev. Lett.} \textbf{\bibinfo{volume}{89}},
  \bibinfo{pages}{220403} (\bibinfo{year}{2002}).

\bibitem[{\citenamefont{Zvonarev et~al.}(2007)\citenamefont{Zvonarev, Cheianov,
  and Giamarchi}}]{zvonarev_ferrobosons_long}
\bibinfo{author}{\bibfnamefont{M.~B.} \bibnamefont{Zvonarev}},
  \bibinfo{author}{\bibfnamefont{V.~V.} \bibnamefont{Cheianov}},
  \bibnamefont{and}
  \bibinfo{author}{\bibfnamefont{T.}~\bibnamefont{Giamarchi}},
  \bibinfo{journal}{in preparation}  (\bibinfo{year}{2007}).

\bibitem[{\citenamefont{Gaudin}(1983)}]{gaudin_book}
\bibinfo{author}{\bibfnamefont{M.}~\bibnamefont{Gaudin}},
  \emph{\bibinfo{title}{La fonction d'onde de Bethe}}
  (\bibinfo{publisher}{Masson}, \bibinfo{address}{Paris},
  \bibinfo{year}{1983}).

\bibitem[{\citenamefont{Fuchs et~al.}(2005)\citenamefont{Fuchs, Gangardt,
  Keilmann, and Shlyapnikov}}]{fuchs_spin_waves_1D_bose}
\bibinfo{author}{\bibfnamefont{J.~N.} \bibnamefont{Fuchs} {\it et.~al.}},
  \bibinfo{journal}{Phys. Rev. Lett.} \textbf{\bibinfo{volume}{95}},
  \bibinfo{pages}{150402} (\bibinfo{year}{2005}).

\bibitem[{\citenamefont{Fiete and Balents}(2004)}]{fiete_spin_decoherent}
\bibinfo{author}{\bibfnamefont{G.~A.} \bibnamefont{Fiete}} \bibnamefont{and}
  \bibinfo{author}{\bibfnamefont{L.}~\bibnamefont{Balents}},
  \bibinfo{journal}{Phys. Rev. Lett.} \textbf{\bibinfo{volume}{93}},
  \bibinfo{pages}{226401} (\bibinfo{year}{2004}).

\bibitem[{\citenamefont{Cheianov and
  Zvonarev}(2004{\natexlab{a}})}]{cheianov_spin_decoherent_short}
\bibinfo{author}{\bibfnamefont{V.~V.} \bibnamefont{Cheianov}} \bibnamefont{and}
  \bibinfo{author}{\bibfnamefont{M.~B.} \bibnamefont{Zvonarev}},
  \bibinfo{journal}{Phys. Rev. Lett.} \textbf{\bibinfo{volume}{92}},
  \bibinfo{pages}{176401} (\bibinfo{year}{2004}{\natexlab{a}}).

\bibitem[{\citenamefont{Cheianov and
  Zvonarev}(2004{\natexlab{b}})}]{cheianov_spin_decoherent_long}
\bibinfo{author}{\bibfnamefont{V.~V.} \bibnamefont{Cheianov}} \bibnamefont{and}
  \bibinfo{author}{\bibfnamefont{M.~B.} \bibnamefont{Zvonarev}},
  \bibinfo{journal}{J. Phys. A} \textbf{\bibinfo{volume}{37}},
  \bibinfo{pages}{2261} (\bibinfo{year}{2004}{\natexlab{b}}).

\bibitem[{\citenamefont{Castella and Zotos}(1993)}]{castella_mob_impurity93}
\bibinfo{author}{\bibfnamefont{H.}~\bibnamefont{Castella}} \bibnamefont{and}
  \bibinfo{author}{\bibfnamefont{X.}~\bibnamefont{Zotos}},
  \bibinfo{journal}{Phys. Rev. B} \textbf{\bibinfo{volume}{47}},
  \bibinfo{pages}{16186} (\bibinfo{year}{1993}).

\bibitem[{\citenamefont{Ogawa et~al.}(1992)\citenamefont{Ogawa, Furusaki, and
  Nagaosa}}]{ogawa_mob_impurity92}
\bibinfo{author}{\bibfnamefont{T.}~\bibnamefont{Ogawa}},
  \bibinfo{author}{\bibfnamefont{A.}~\bibnamefont{Furusaki}}, \bibnamefont{and}
  \bibinfo{author}{\bibfnamefont{N.}~\bibnamefont{Nagaosa}},
  \bibinfo{journal}{Phys. Rev. Lett.} \textbf{\bibinfo{volume}{68}},
  \bibinfo{pages}{3638} (\bibinfo{year}{1992}).

\bibitem[{\citenamefont{Tsukamoto et~al.}(1998)\citenamefont{Tsukamoto, Fujii,
  and Kawakami}}]{tsukamoto_mob_impurity98}
\bibinfo{author}{\bibfnamefont{Y.}~\bibnamefont{Tsukamoto}},
  \bibinfo{author}{\bibfnamefont{T.}~\bibnamefont{Fujii}}, \bibnamefont{and}
  \bibinfo{author}{\bibfnamefont{N.}~\bibnamefont{Kawakami}},
  \bibinfo{journal}{Phys. Rev. B} \textbf{\bibinfo{volume}{58}},
  \bibinfo{pages}{3633} (\bibinfo{year}{1998}).

\bibitem[{\citenamefont{Sorella and Parola}(1996)}]{sorella_hubbmott_mobhole96}
\bibinfo{author}{\bibfnamefont{S.}~\bibnamefont{Sorella}} \bibnamefont{and}
  \bibinfo{author}{\bibfnamefont{A.}~\bibnamefont{Parola}},
  \bibinfo{journal}{Phys. Rev. Lett.} \textbf{\bibinfo{volume}{76}},
  \bibinfo{pages}{4604} (\bibinfo{year}{1996}).

\bibitem[{\citenamefont{Caldeira and Leggett}(1983)}]{caldeira_leggett}
\bibinfo{author}{\bibfnamefont{A.~O.} \bibnamefont{Caldeira}} \bibnamefont{and}
  \bibinfo{author}{\bibfnamefont{A.~J.} \bibnamefont{Leggett}},
  \bibinfo{journal}{Ann. Phys.} \textbf{\bibinfo{volume}{149}},
  \bibinfo{pages}{374} (\bibinfo{year}{1983}).

\bibitem[{\citenamefont{Hakim and Ambegaokar}(1985)}]{hakim_dissip85}
\bibinfo{author}{\bibfnamefont{V.}~\bibnamefont{Hakim}} \bibnamefont{and}
  \bibinfo{author}{\bibfnamefont{V.}~\bibnamefont{Ambegaokar}},
  \bibinfo{journal}{Phys. Rev. A} \textbf{\bibinfo{volume}{32}},
  \bibinfo{pages}{423} (\bibinfo{year}{1985}).

\bibitem[{\citenamefont{Lieb and
  Mattis}(1962)}]{lieb_mattis_fermions_unmagnetized}
\bibinfo{author}{\bibfnamefont{E.}~\bibnamefont{Lieb}} \bibnamefont{and}
  \bibinfo{author}{\bibfnamefont{D.}~\bibnamefont{Mattis}},
  \bibinfo{journal}{Phys. Rev.} \textbf{\bibinfo{volume}{125}},
  \bibinfo{pages}{164} (\bibinfo{year}{1962}).

\end{thebibliography}
\end{document}